\documentclass[a4paper,11pt]{article}
\usepackage{tikz}
\usetikzlibrary{positioning, arrows.meta}
\usepackage{siunitx}	
\usepackage{pos}
\usepackage{booktabs}
\usepackage{float}
\usepackage[symbol]{footmisc}
\usepackage{cleveref}
\usepackage{multirow}
\usepackage{printlen}
\usepackage{makecell}
\usepackage{tikz}

\newcommand{\vB}{v \times B}
\newcommand{\vvB}{v \times v \times B}
\newcommand{\xmax}{X_\mathrm{max}}

\title{Generative Neural Network for Simulating Radio Emission from Extensive Air Showers}
 \ShortTitle{Neural Network for Radio Simulations}

\author*[a]{Pranav Sampathkumar}
\author[a,b]{Tim Huege}
\author[a]{Andreas Haungs}
\author[a]{Ralph Engel}

\affiliation[a]{Institute for Astroparticle Physics, Karlsruhe Institute of Technology,\\
Karlsruhe, Germany}
\affiliation[b]{Vrije Universiteit Brussel, Astrophysical Institute, Pleinlaan~2, 1050 Brussels, Belgium}

\emailAdd{pranav.sampathkumar@kit.edu}

\abstract{
Cosmic ray shower detection using large radio arrays has gained significant traction in recent years.
With massive improvements in signal modelling and microscopic simulations, the analysis of incoming events is still severely limited by the simulation cost of radio emission to interpret the data.
In this work, we show that a neural network can be used for simulating such radio pulses.
We also demonstrate how such a neural network can be used for $\xmax$ reconstruction, while retaining comparable resolution to using full Monte-Carlo CORSIKA/CoREAS simulations for radio emission. 
}

\ConferenceLogo{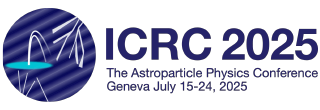}
\FullConference{39th International Cosmic Ray Conference (ICRC2025)\\
 15–24 July 2025\\
Geneva, Switzerland\\}

\begin{document}
\maketitle
\section{Introduction}
The use of radio emission from extensive air showers (EAS) has gained significant traction in the past decade with various advances to signal processing and instrumentation using massive arrays of antennas.
These advances have led to reliable reconstruction approaches for cosmic ray events using radio antennas.
These reconstructions provide valuable insights into the energy spectrum and mass composition of cosmic rays, while radio arrays are cost-efficient for deployment over vast areas.
Various current/planned experiments such as AERA\cite{aera}, LOFAR \cite{lofar}, SKA \cite{SKA} IceCube-Gen2 \cite{icecube-gen2}, RNO-G \cite{rnog} and GRAND \cite{grand} etc., utilize radio emission in order to improve the study of cosmic rays. 
The current reconstruction methodology used for these arrays involves microscopic simulations of the electromagnetic emission from individual tracks of the particles.
This makes these simulations computationally heavy as the individual contributions from all these tracks need to be calculated and added up for every antenna.
In some established analysis approaches \cite{aera_xmax_paper, lofar_xmax_paper}, multiple iterations of each measured event need to be simulated in order to perform a detailed study, while keeping track of the uncertainties arising in the radio measurements, which are noisy in nature.
This makes the computational cost a massive bottleneck, in the reconstruction of new radio data at higher and higher energies. 

Previous works have been done in order to reduce this computational cost for big arrays of antennas. A neural network to simulate the longitudinal profile quickly without explicit Monte Carlo was developed \cite{my_icrc_2023}. For radio simulations, A template based approach using a single simulated shower utilizing semi-analytical relationships could drastically reduce simulation times for multiple iterations \cite{template_synthesis}. 
Also, a pulse interpolation method was developed as to not perform the simulation for every antenna position \cite{pulse_interpolation}.
The method still relies on conventional simulations for the star shaped antenna positions from which the other positions are interpolated.
In this work, we provide with a neural network which can simulate radio emission for a given position in a wide variety of event geometries.
We will then show that this neural network can be used for reconstruction via the shower maximum $X_{max}$, for individual events, thus reducing the simulation times drastically for estimating the primary mass from radio detection.

\section{Training Data}
\begin{figure*}[ht]
    \centering
    \includegraphics[width=0.32\linewidth]{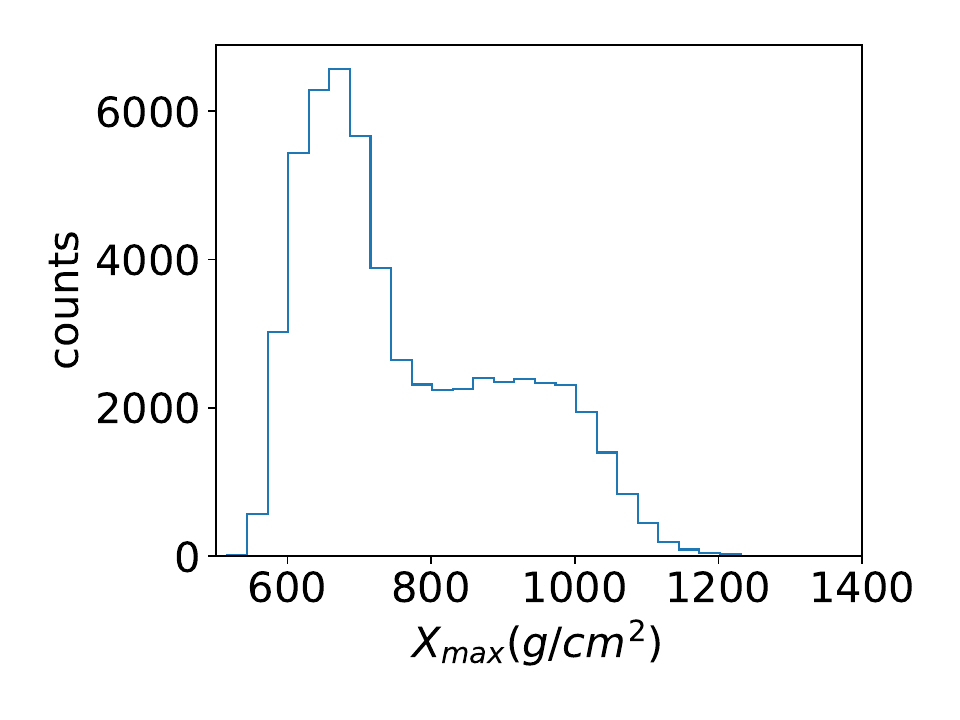}
    \includegraphics[width=0.32\linewidth]{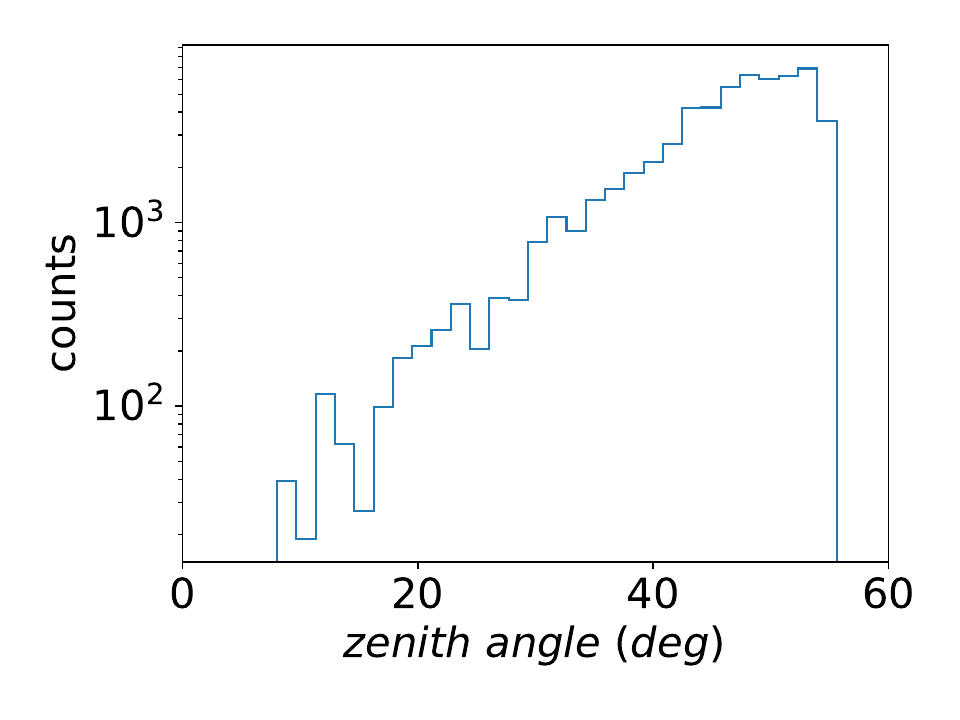}
    \includegraphics[width=0.32\linewidth]{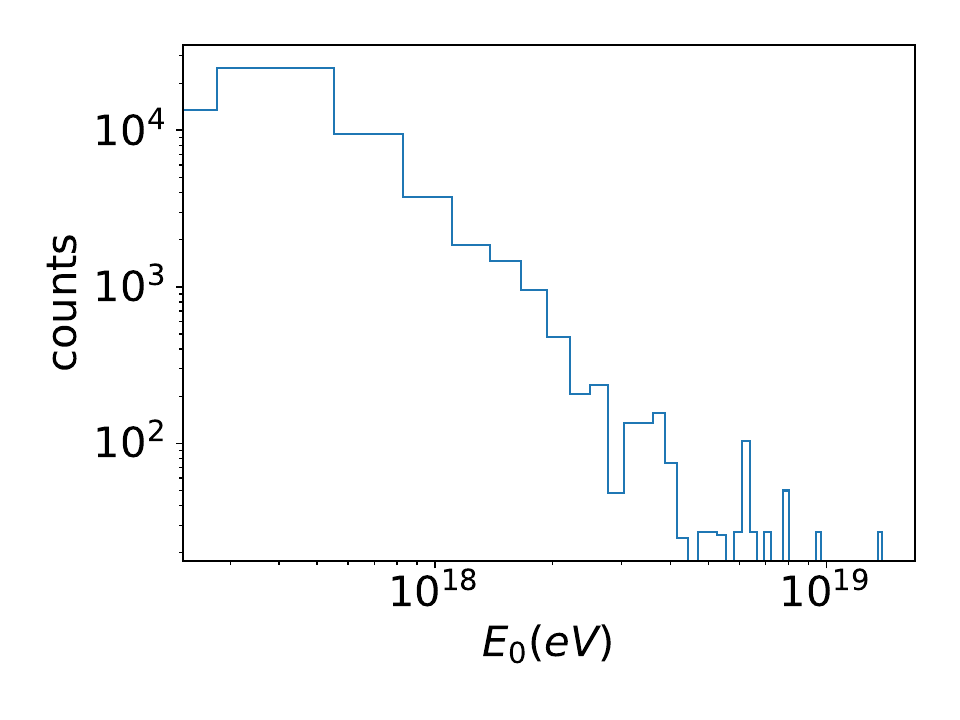}
    \caption{Shown are the distributions of $\xmax$, zenith angle $\theta$, and primary energy $E_0$ for the full CORSIKA/CoREAS simulations. These show the phase space distribution of the training data and show us the regions the network is trained on. It is expected that the network works well in this phase-space region.}
    \label{fig:input_dist}
\end{figure*}

As with any machine learning algorithm, we need a high quality training dataset in order to train our neural network. 
For this work, we are using the vast library of CoREAS \cite{coreas} simulations built for analysing events within AERA (Auger Engineering Radio Array) \cite{aera_xmax_paper}\footnote{We are grateful to the Pierre Auger Collaboration for allowing us to use their simulation set within this work.}.
These simulated events were specifically made with the time-variable atmosphere above AERA, and thus the neural network trained in this work, can only be used in the context of this experiment. 
The network is also trained in AERA's frequency band, 30-80MHz. 
Since the distribution in phase space for these simulations is realistic in terms of the distribution of measured showers by AERA, the model is expected to lose fidelity in rare events. 
The distribution of events is shown in \Cref{fig:input_dist}.
These simulations were set up with CORSIKA 7.7100 \cite{corsika7} with radio emission from particle tracks done with the CoREAS extension \cite{coreas}.
QGSJETII-04 is the hadronic interaction model in these simulations \cite{qgsjet2}.
The atmospheric model is using the \textit{Global Data Assimilation System} (GDAS) \cite{gdas} via \textit{gdastool} \cite{gdas-tool}.
The resulting library has 2158 different shower paramters with 27 different iterations of each of these showers, resulting in a total of around 58k simulations.
This work serves as a proof of concept within the context of AERA
\footnote{
However, the model can be fine-tuned with a smaller dataset for changes in experimental setup and frequency bands.  Such a model has been successfully trained for LOFAR simulations \cite{denis_lofar}.
}, and we hope that in the future,  if neural networks are part of the
simulation pipeline in experiments, a dedicated simulation training set can be made to solve the loss of fidelity in rare events. 
We use the star shaped antenna positions in each of these simulations (240 antennas). Of this dataset, 80\% of the simulations are used for training and the remaining 20\% are used for testing purposes. The antenna position and pulses are transformed to the shower plane. Also, the weak pulses, which are dominated by thinning artifacts during simulations, are removed.
\section{Neural Network Model}
\label{sec:nn_model}
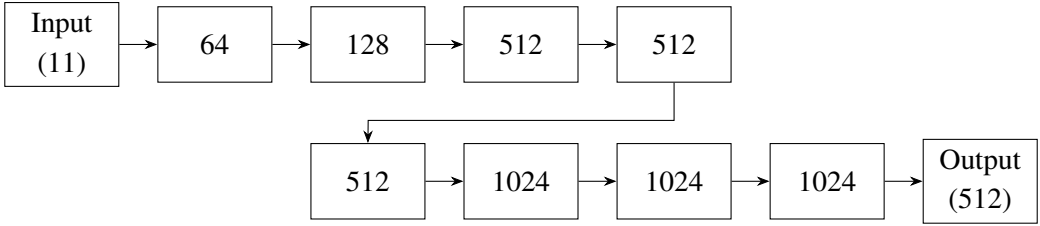
\begin{figure}[h]
    \centering
\begin{tikzpicture}[
    node distance=0.8cm and 0.5cm,
    every node/.style={draw, minimum height=1cm, minimum width=1.5cm, align=center},
    ->, >=Stealth
]

\node (input) {Input\\(11)};
\node (l1) [right=of input] {64};
\node (l2) [right=of l1] {128};
\node (l3) [right=of l2] {512};
\node (l4) [right=of l3] {512};
\node (l5) [below=of l2] {512};
\node (l6) [right=of l5] {1024};
\node (l7) [right=of l6] {1024};
\node (l8) [right=of l7] {1024};
\node (output) [right=of l8] {Output\\(512)};

\draw[->] (input) -- (l1);
\draw[->] (l1) -- (l2);
\draw[->] (l2) -- (l3);
\draw[->] (l3) -- (l4);
\draw[->] (l4.south) -- ++(0,-0.5) -- ++(-4.025,0) -- (l5.north);
\draw[->] (l5) -- (l6);
\draw[->] (l6) -- (l7);
\draw[->] (l7) -- (l8);
\draw[->] (l8) -- (output);

\end{tikzpicture}
    \caption{Neural network schematic for generating radio pulses:
    The 11 inputs are described in \Cref{sec:nn_model}, and the 512 outputs are the 256 time bins in the 2 polarisations.
    The network produces a pulse, given an antenna position and shower parameters.
    The model has about 4M parameters, and has a memory footprint of 19MB. 
    }
    \label{fig:nn_schematic}
\end{figure}

Neural networks have been gaining substantial interest as an alternative simulation tool for physics. 
The rise of generative AI techniques has helped speed up physics simulations significantly.
We use a fully connected neural network for our case here, which is being directly trained to generate the pulses at each antenna position. 
This network uses the atmospheric depth at which the particle count reaches it's maximum ($\xmax$), the electromagnetic energy, geomagnetic angle, density and height at $\xmax$, primary energy of the incoming cosmic ray, arrival direction of the shower via zenith and azimuth angles and the antenna position in shower coordinates as input. 
We have a total of 11 input parameters, and the network predicts pulses in the $\vB$ and $\vvB$ polarisations, each. 
The pulses are predicted with 256 time bins in each polarisation with a timing resolution of 1\,ns, resulting in 512 output nodes for the network.
The network has 8 hidden layers, and the schematic is given in \Cref{fig:nn_schematic}.
There are no skip connections, as it was found that the gradients do not vanish at this depth.
We scaled the input values by nominal values so that the distribution lies between -1 and 1. 
The training process has an L1 norm ($\mathcal{L} = | x- y|$) as the loss function, and we weight the $\vvB$ polarisation more, since the dominant emission is within the $\vB$ polarisation.
Weighting the weaker polarisation accordingly makes sure that both polarisations are learnt correctly.
The L1 norm was chosen so that the weaker pulses, whose contributions to the loss will be low, are still represented sufficiently.
The minimisation algorithm used is ADAM \cite{adam}, and regularisation is done via weight decay during minimisation.
The pipeline is implemented using Pytorch and Numpy \cite{pytorch,numpy}.
The training procedure is fast and can be completed within a week in a local desktop computer on CPUs \footnote{Tested on a AMD Ryzen 7 PRO 3700 8-Core Processor, 3.6 GHz}.
After training, inference requires only a few hundred milliseconds for each antenna position simulated.
Neural networks also have the advantage of being optimized to use GPUs and multi-threaded CPUs. Thus, this helps parallelise the radio-emission simulations, which is an ongoing effort by the community \cite{c8_coreas_parallel}.

\section{Performance}
\subsection{Pulses}
\begin{figure*}[!htb]
    \centering
    \begin{tabular}{m{0.15\linewidth}|>{\centering}m{0.37\linewidth}|>{\centering\arraybackslash}m{0.37\linewidth}}
    & CoREAS & Neural Network \\
    \hline
    \makecell{$\vB$\\ polarization} & 
    \includegraphics[width=\linewidth]{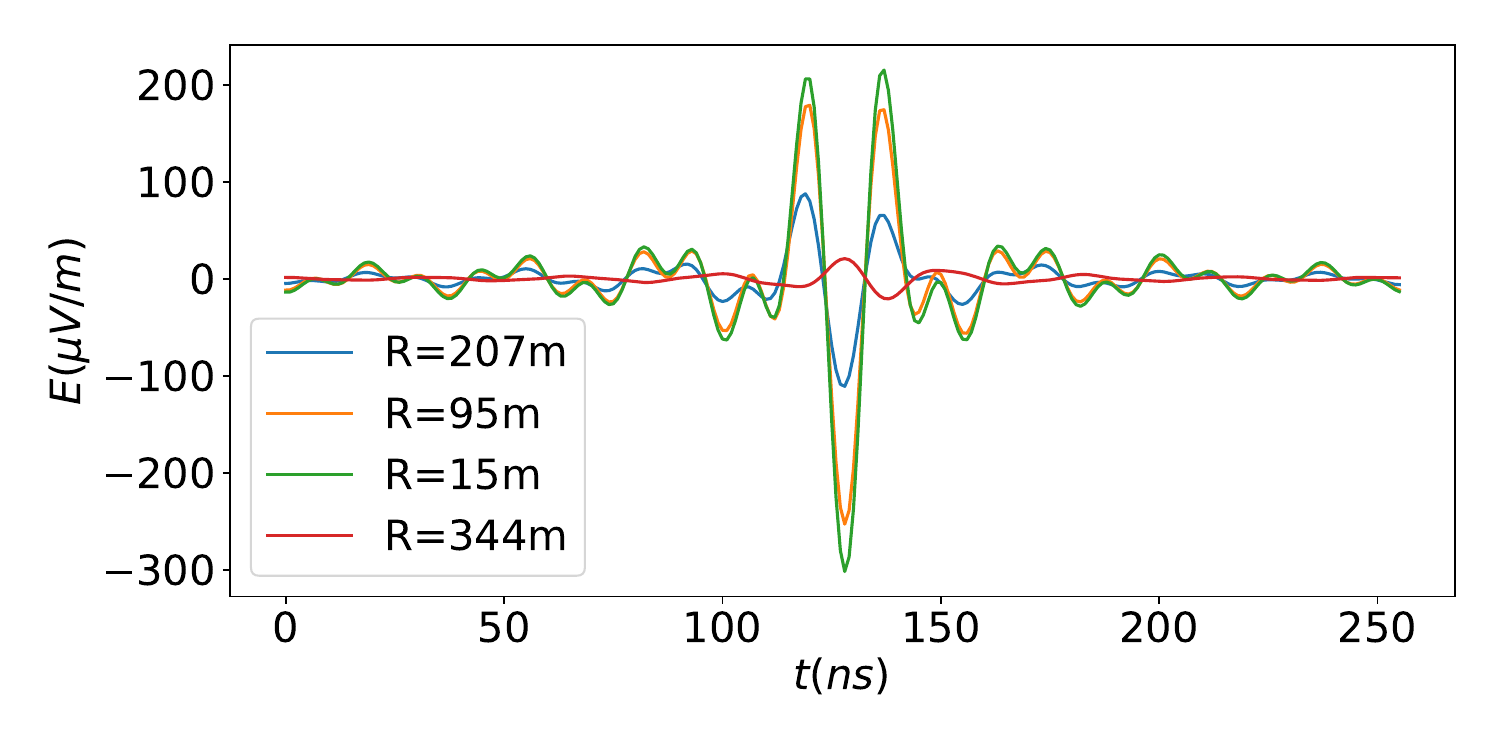} &
    \includegraphics[width=\linewidth]{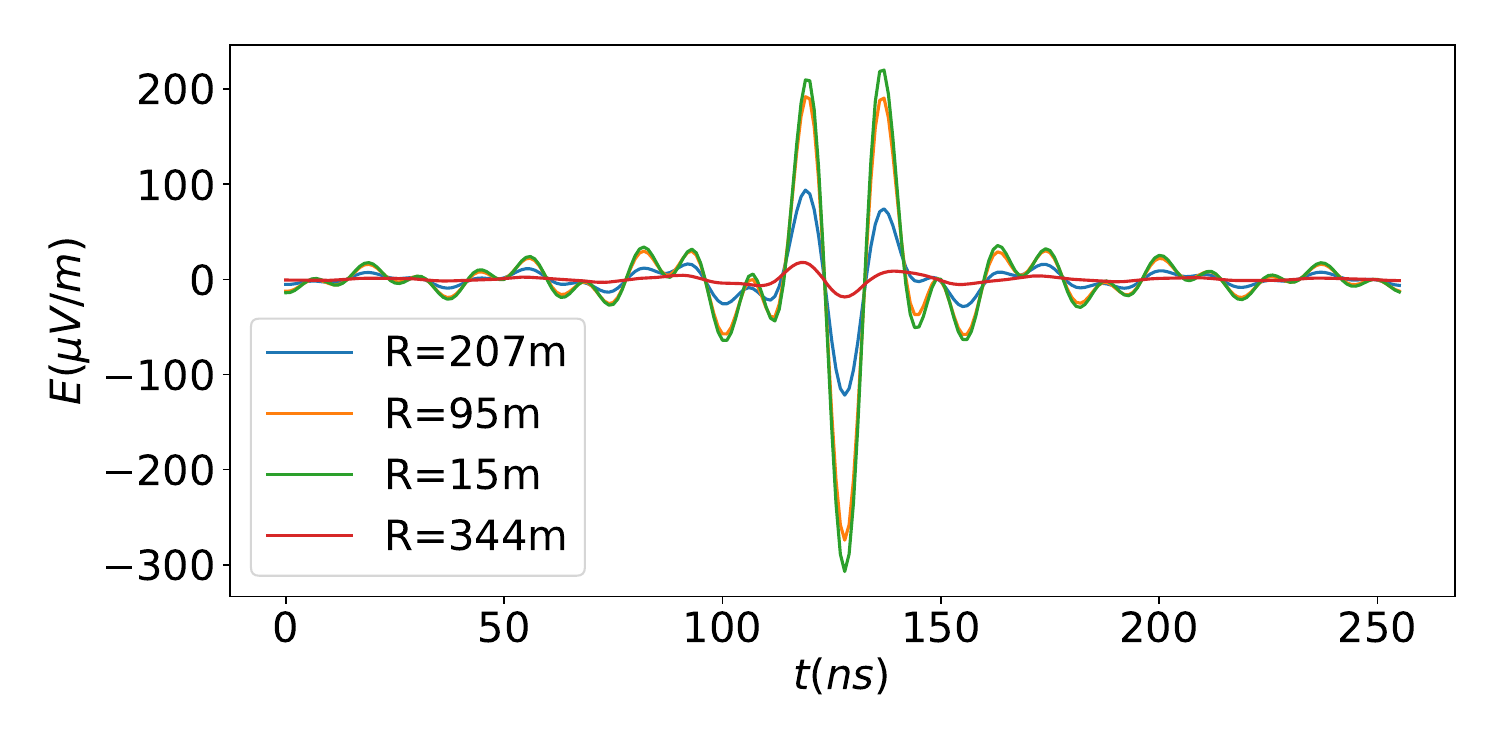}\\
    \hline
    \makecell{$\vvB$\\ polarization} & 
    \includegraphics[width=\linewidth]{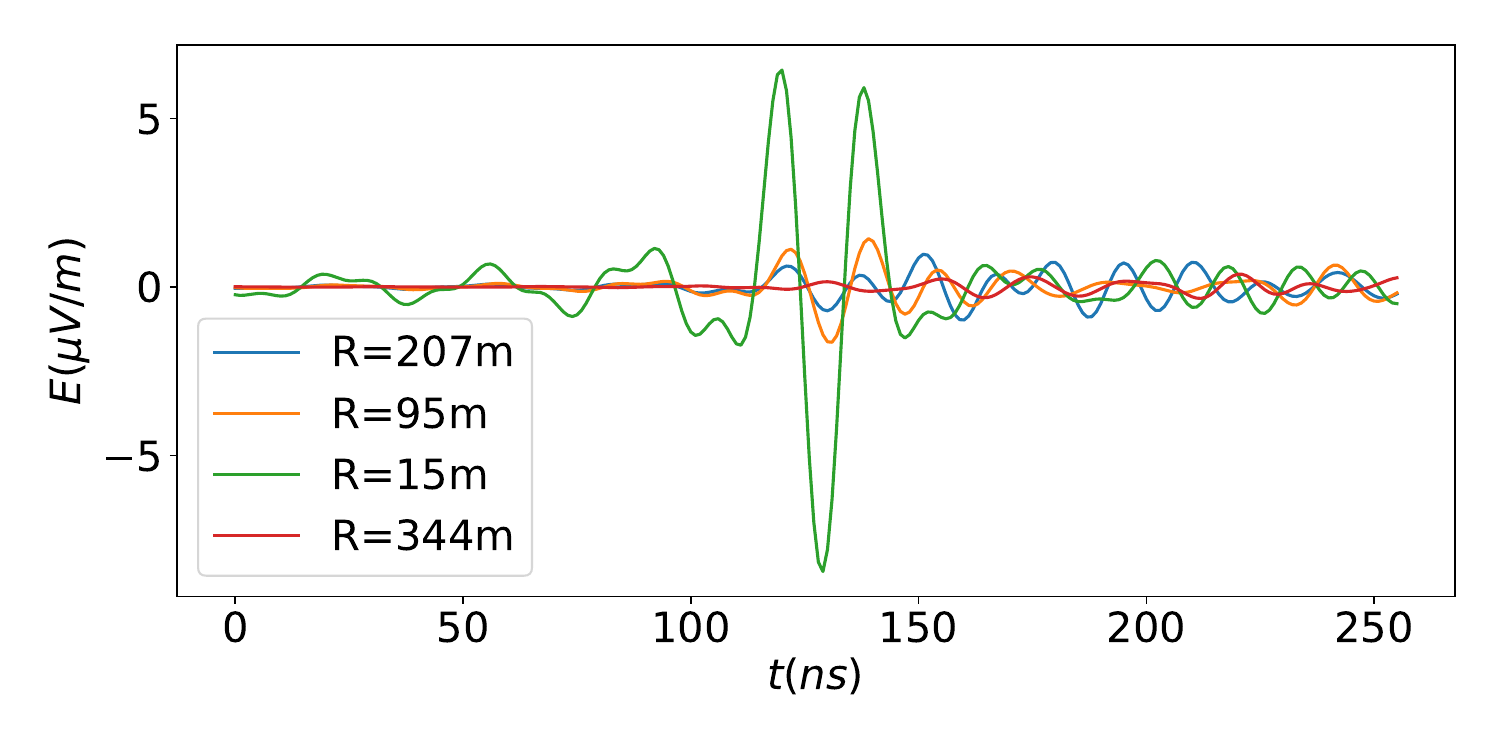} &
    \includegraphics[width=\linewidth]{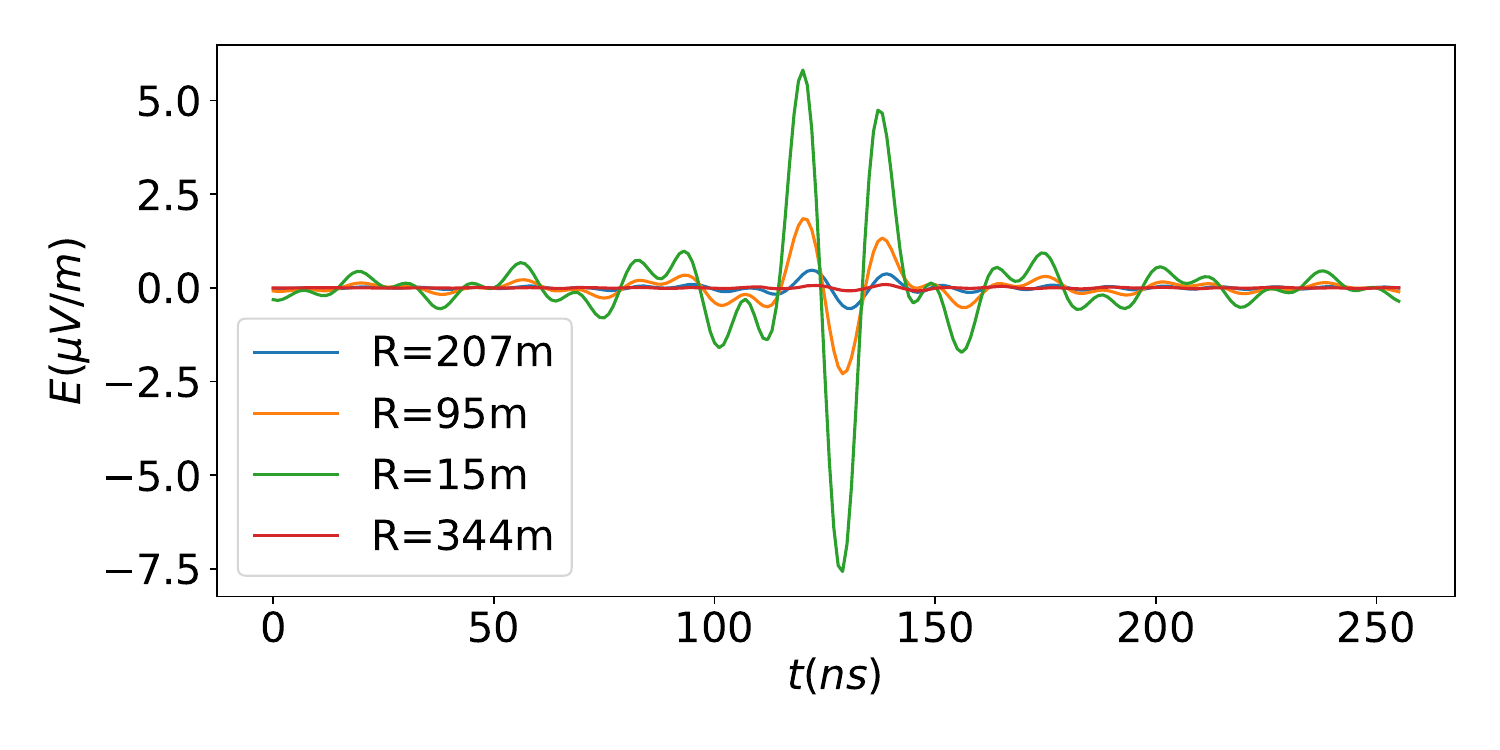}\\
    \end{tabular}
    \caption{Comparison of generated radio pulses between a CoREAS simulation and neural network-based simulation for both polarizations at antennas at various distances from the shower core (R). 
    This is a simulation with $\theta$ = \ang{43.56} and primary energy of 
    \qty{3e8}{\giga\electronvolt}  with an $\xmax$ of 
    \qty{703}{\gram\per\square\centi\metre}. 
    We see the agreement is quite good for strong pulses. A more quantitative comparison between the pulses is shown in \Cref{fig:corr_vs_flu}.
    }
    \label{fig:pulses}
\end{figure*}

The neural network can directly produce centered pulses in the trained time resolution. We can see that the pulses agree qualitatively. We study the variation with pulses with varying geometries and $\xmax$ to see the agreement, of which one example is shown in \Cref{fig:pulses}
\clearpage
\subsection{Fluence Error and Correlation between Pulses}
\begin{figure*}[!htb]
    \centering
    \begin{tabular}{m{0.15\linewidth}|>{\centering}m{0.37\linewidth}|>{\centering\arraybackslash}m{0.37\linewidth}}
    & CoREAS & Neural Network \\
    \hline
    \makecell{$\vB$\\ polarization} & \includegraphics[width=\linewidth]{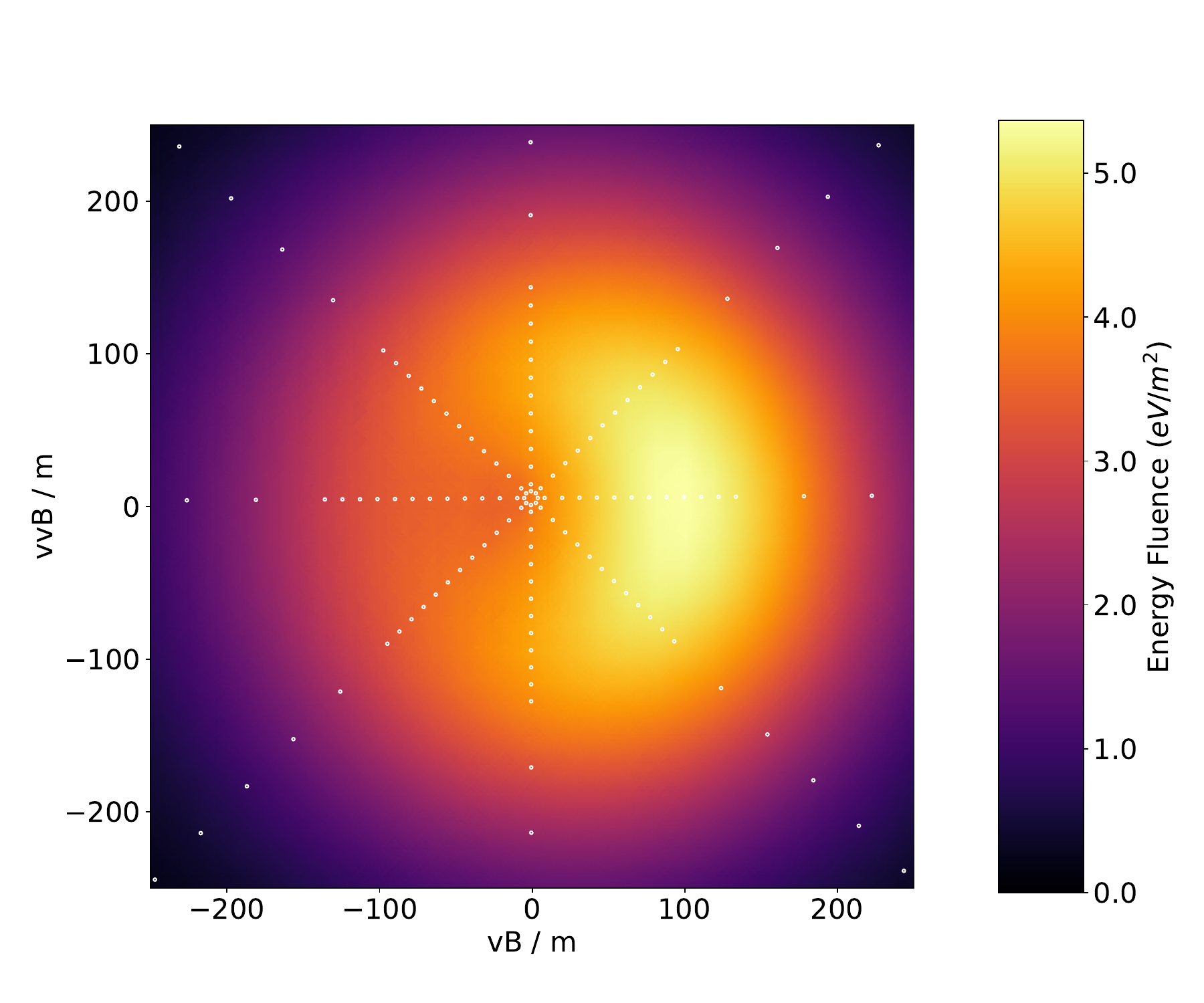} &
    \includegraphics[width=\linewidth]{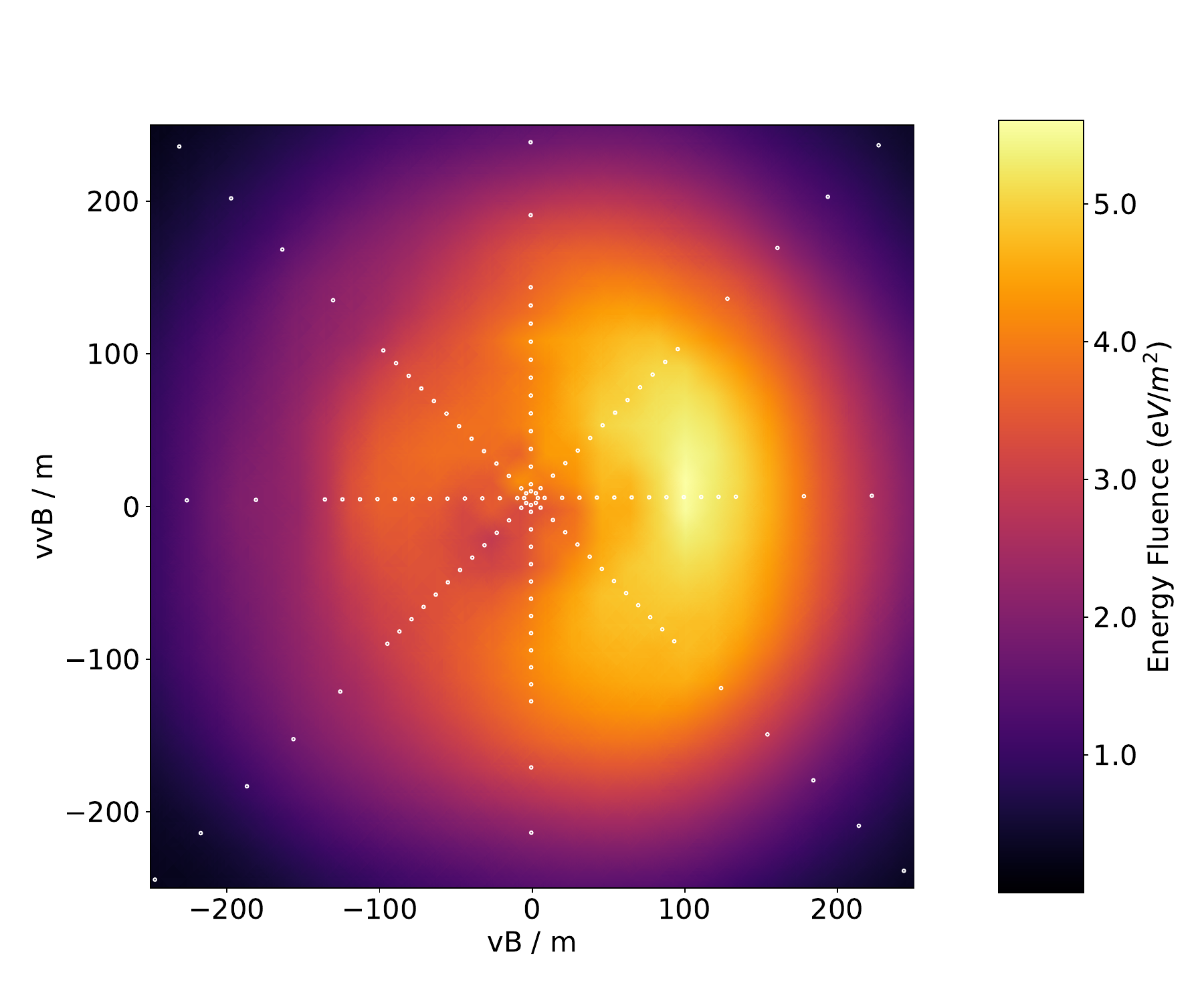}\\
    \hline
    \makecell{$\vvB$\\ polarization} & \includegraphics[width=\linewidth]{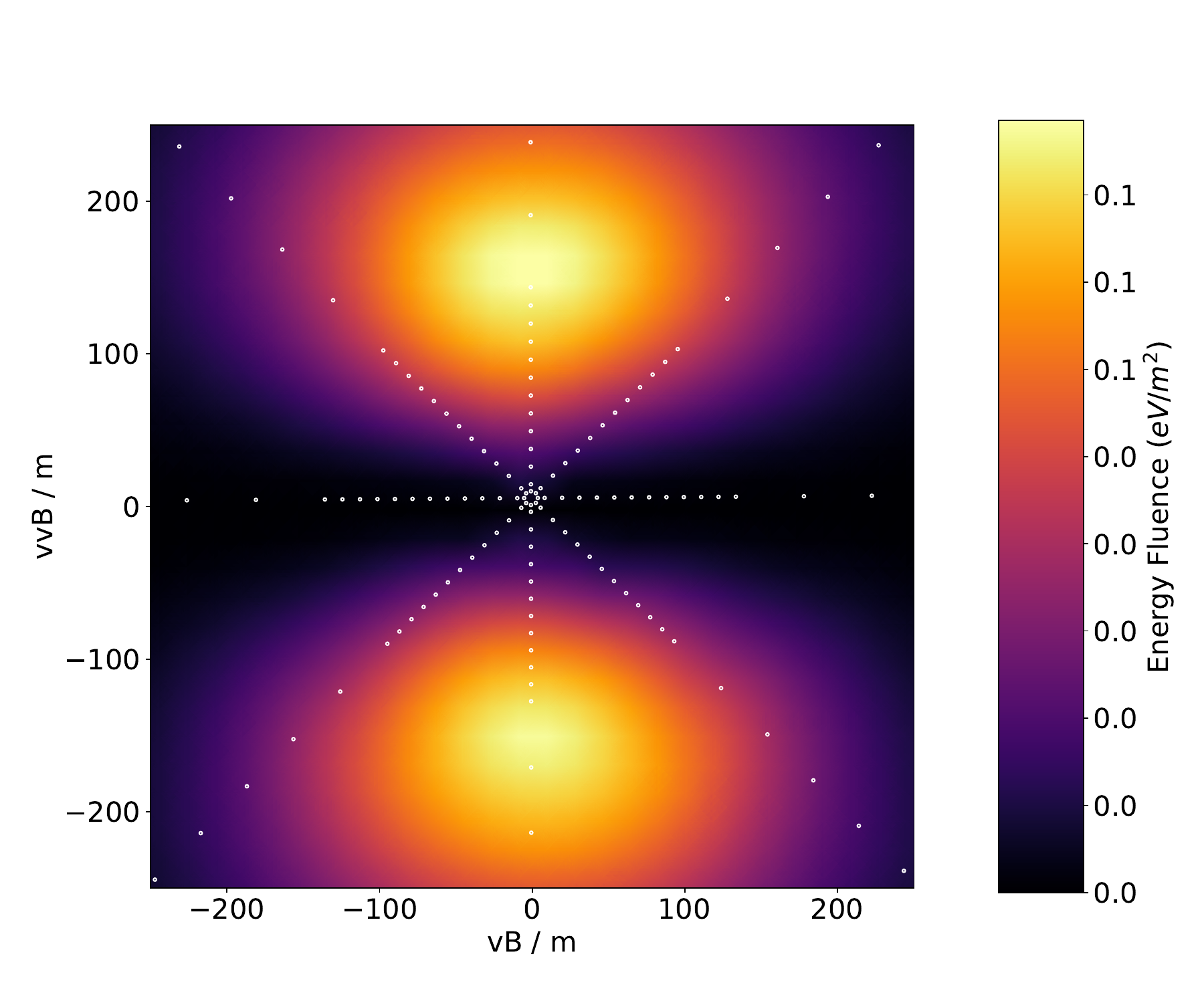} & 
    \includegraphics[width=\linewidth]{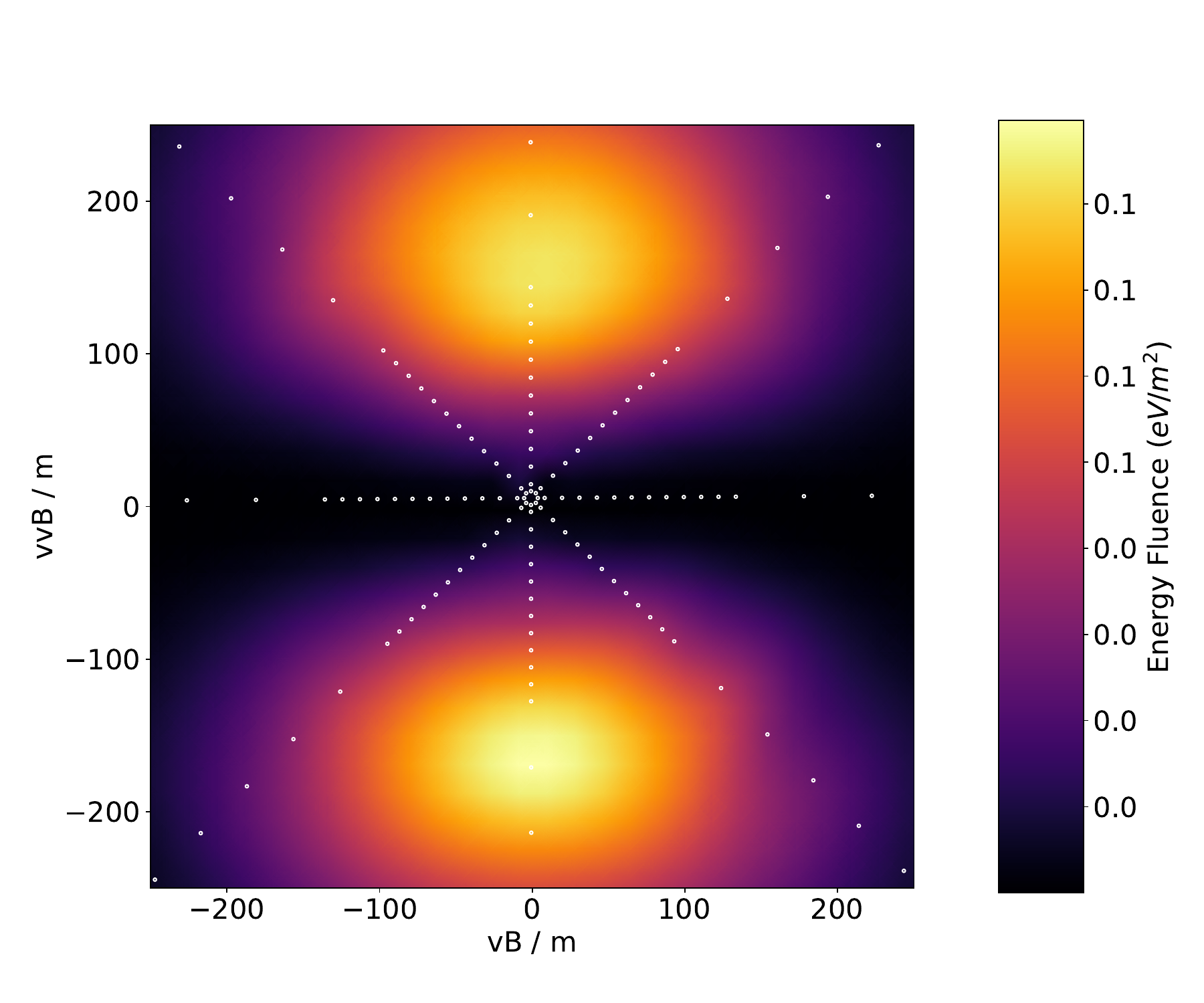}\\
    \end{tabular}
    \caption{Comparison of fluence maps between a CoREAS simulation and neural network-based simulation for both polarizations. 
    This is a simulation with $\theta$ = \ang{54.1} and primary energy of 
    \qty{2.94e9}{\giga\electronvolt}  with an $\xmax$ of 
    \qty{733}{\gram\per\square\centi\metre}. The pulses are simulated at the indicated antenna positions, afterwards the intermediate values are calculated by radially interpolating the fluence using the \texttt{scipy.Rbf} module. 
    }
    \label{fig:flu_comp}
\end{figure*}
\begin{figure}[bt]
    \centering
    \includegraphics[width=0.45\textwidth]{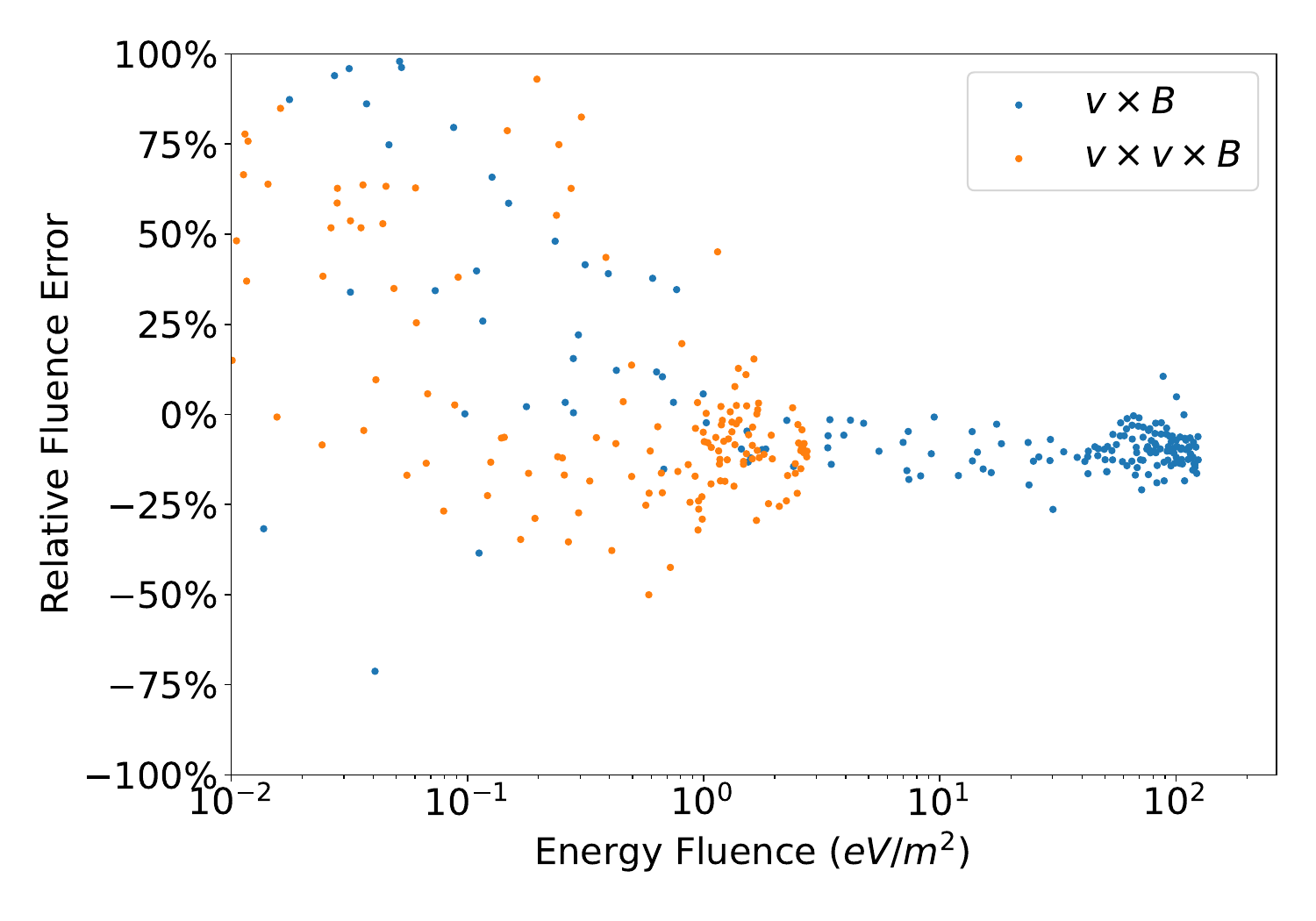}
    \includegraphics[width=0.45\textwidth]{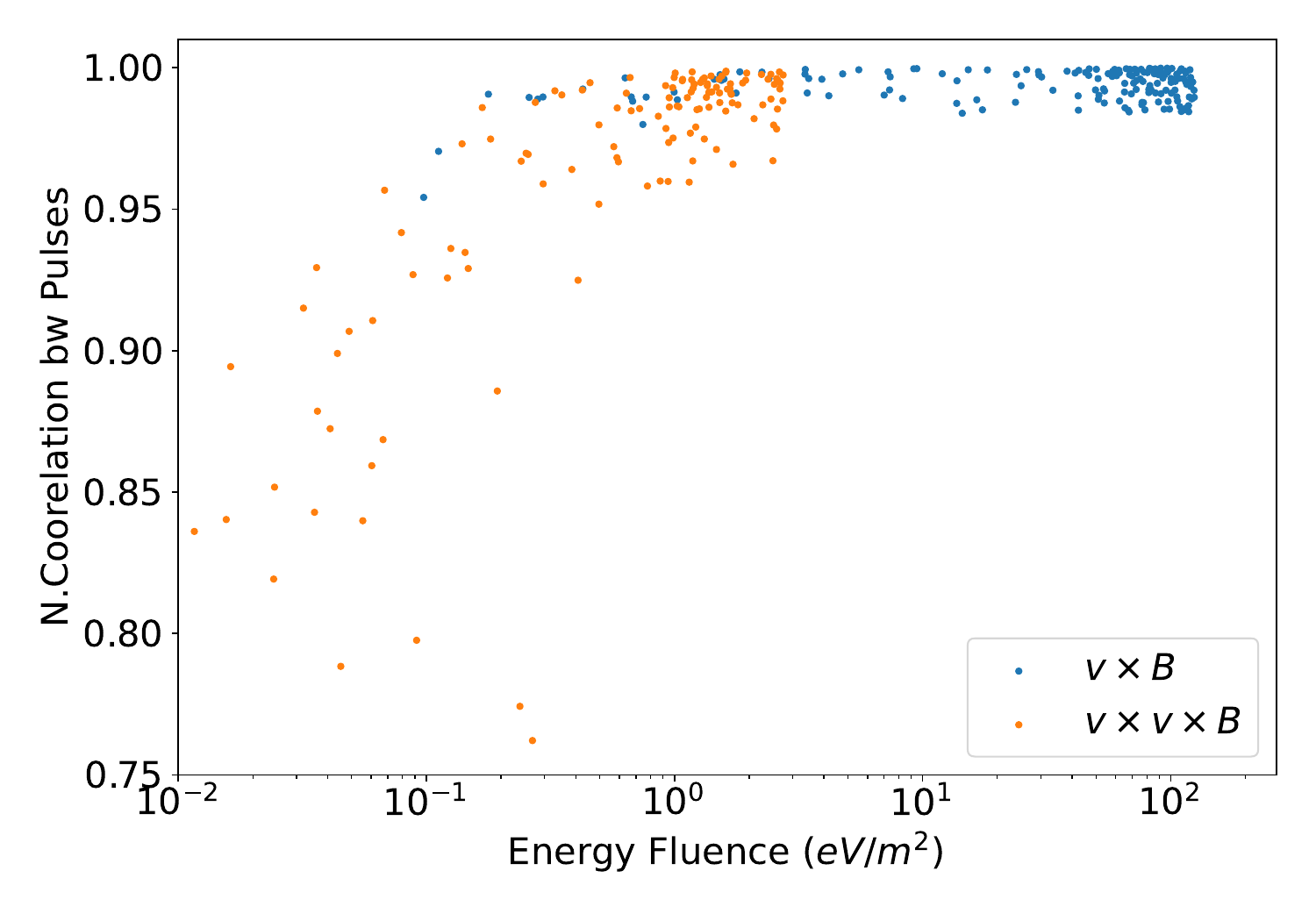}
    \caption{The relative fluence error and the normalized correlation between the pulses simulated using CoREAS and Neural Network for 240 antennas in a star shaped grid for a typical shower are given here. We can see that the agreement is quite high for high fluence pulses, which would contribute to the measured pulses with a high SNR value.
    This is a simulation with $\theta$ = \ang{44.82} and primary energy of 
    \qty{8.85e8}{\giga\electronvolt}  with an $\xmax$ of 
    \qty{718}{\gram\per\square\centi\metre}. 
    }
    \label{fig:corr_vs_flu}
\end{figure}

The energy deposited per unit area (the fluence) is critical in reconstructing shower parameters based on measured signals. In \Cref{fig:flu_comp} we see the comparison of fluences between the CoREAS radio pulses and the radio pulses provided by the neural network. The fluences are found to mostly match within 10\% accuracy, and thus the network shows promise and can be used for $\xmax$ reconstruction procedure. 
In addition, a quantitative comparison of pulses between the neural network and the CoREAS simulations is important since the pulse shape contain valuable information regarding shower development \cite{pulse_shape}. We compare the fluence and correlation between the pulses at an antenna level in \Cref{fig:corr_vs_flu}, which show good agreement between the pulses. 
\subsection{$\xmax$ reconstruction}
In order to check the validity of the network for an $\xmax$
reconstruction, we do a simplistic $\xmax$ reconstruction procedure which is inspired by the fluence-based method used by many experiments \cite{aera_xmax_paper, lofar_xmax_paper}.
These approaches work by calculating the fluence at each antenna and then comparing it with the fluence of various simulated events.
The $\chi^2$ difference between the measured and simulated fluences is calculated with varying $\xmax$, and following which the point of minimum $\chi^2$ is interpolated and taken as the reconstructed $\xmax$.

The $\chi^2$ difference between the measured shower and simulations is defined as
\begin{align}
    \label{eq:chi2}
    \chi^2 = \sum_\mathrm{antennas} \left( \frac{P_\mathrm{ant} - S^2 P_{sim}(x_\mathrm{ant} - x_0, y_\mathrm{ant} - y_0)}{\sigma_\mathrm{ant}} \right)^2
\end{align}
where S is a scaling factor which accounts for multiplicative deviations between fluence measurements and simulations. $x_0$ and $y_0$ denote the core position and $\sigma_{ant}$ denotes the uncertainties in fluence measurements at every antenna. For our purposes, since we are comparing simulations with simulations, we set $S=1$  and assume a particular model, described below, for the uncertainties. These calculations are also done for simulations where the core is at the origin of the coordinate system, i.e., we we do not vary the core position during the fitting procedure, as needed in measurements.

\begin{figure}[ht]
    \centering
    \includegraphics[width=0.49\linewidth]{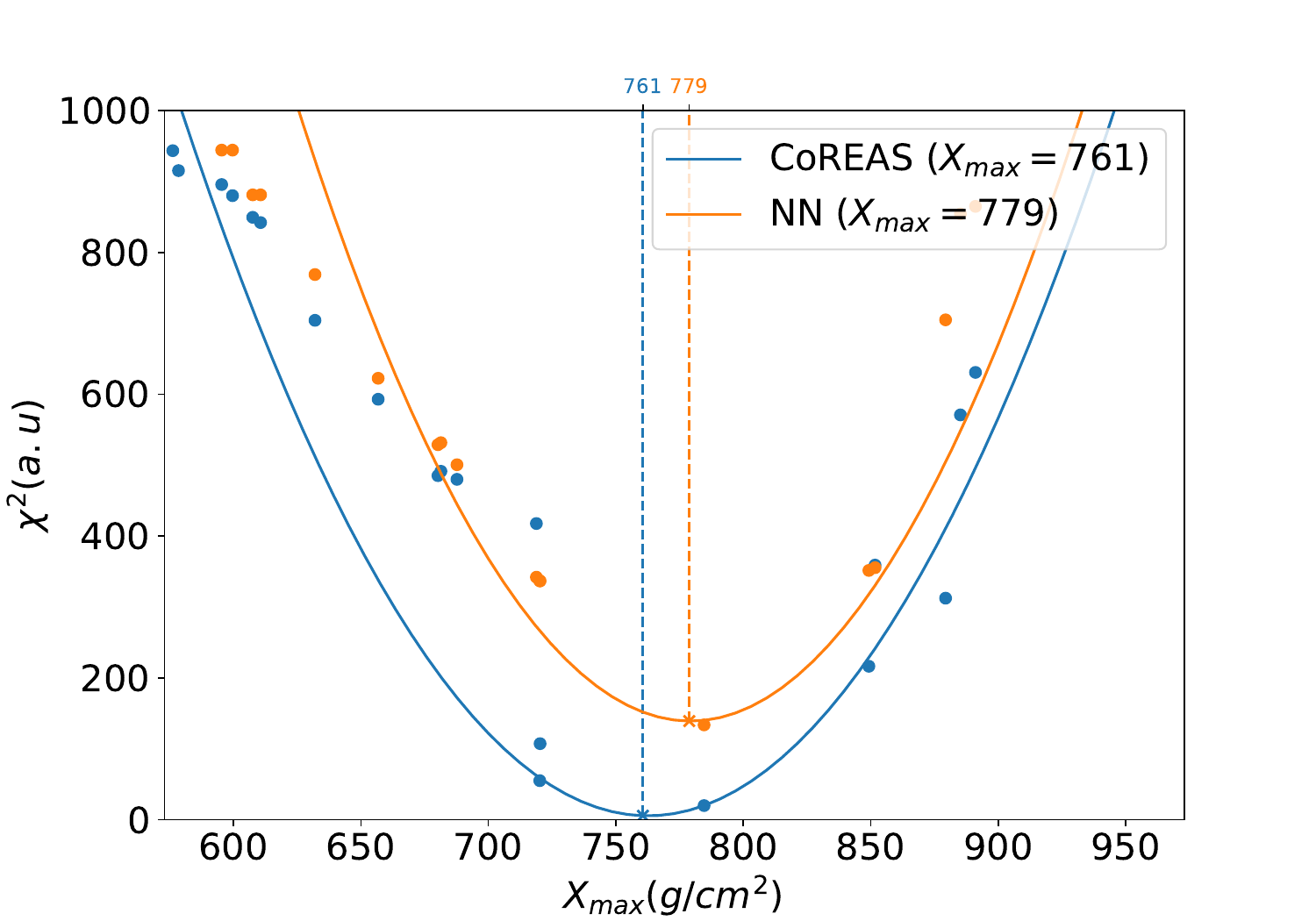}
    \includegraphics[width=0.49\linewidth]{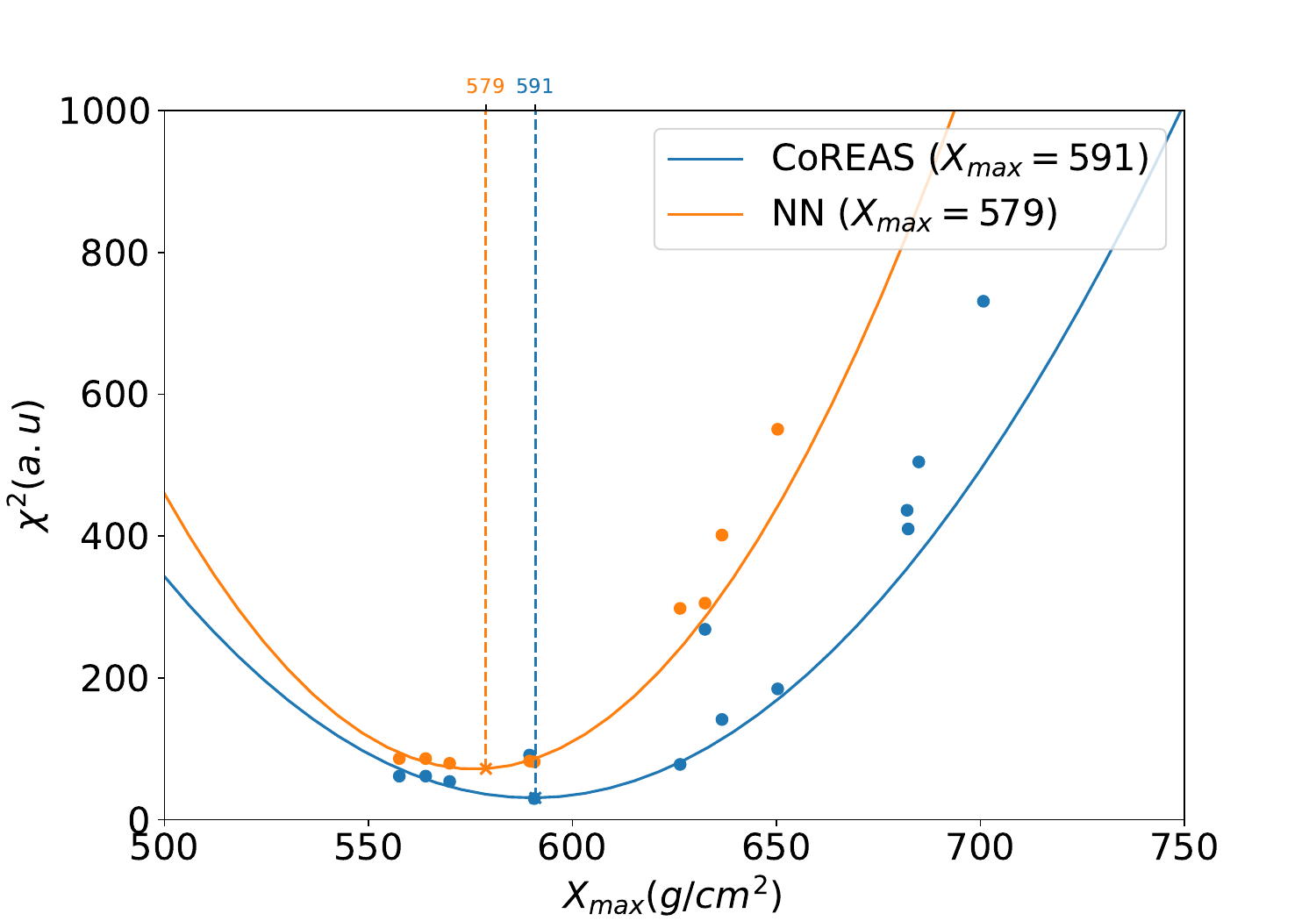}
    \caption{We show the $\xmax$ reconstruction for two events using a fluence based $\chi^2$ method. The same points which are used for the traditional reconstruction, are also sampled using a neural network and a parabola is fit.
    This highlights that the network has a correlation between $\xmax$ and pulses, and the network can be used for the reconstruction procedure. In the above figure, the left plot is from an event with true $\xmax$ of \qty{773}{\gram\per\square\centi\metre} and the right plot has a true $\xmax$ of \qty{585.4}{\gram\per\square\centi\metre}.}
    \label{fig:xmax_reco_coreas}
\end{figure}

For the noise model, we assume a Gaussian multiplicative noise (representing antenna-to-antenna gain variations) and an additive noise to add to the signals. The additive noise is 1 $\mu V/m^2$ to the fluences. The method then involves freely simulating around the initial estimate of $\xmax$. A parabola is then fit to the $\chi^2$ values and the minimum of the parabola is used for the reconstructed $\xmax$ value. This method for two different showers as shown in \Cref{fig:xmax_reco_coreas}.
Certain properties common in both the simulation methods, are noticeable.
The build-up of bias at low $\xmax$ is due to the parabola fitting procedure in the $\xmax$ reconstruction method.
Since the reconstruction method involves limited sampling, the parabola fit becomes biased with no equal number of simulations on either side of the parabola, leading to a bias in lower $\xmax$.

\subsection{Bias and Resolution of $\xmax$ reconstruction}
\begin{figure*}[t]
    \centering
    \includegraphics[width=0.45\textwidth]{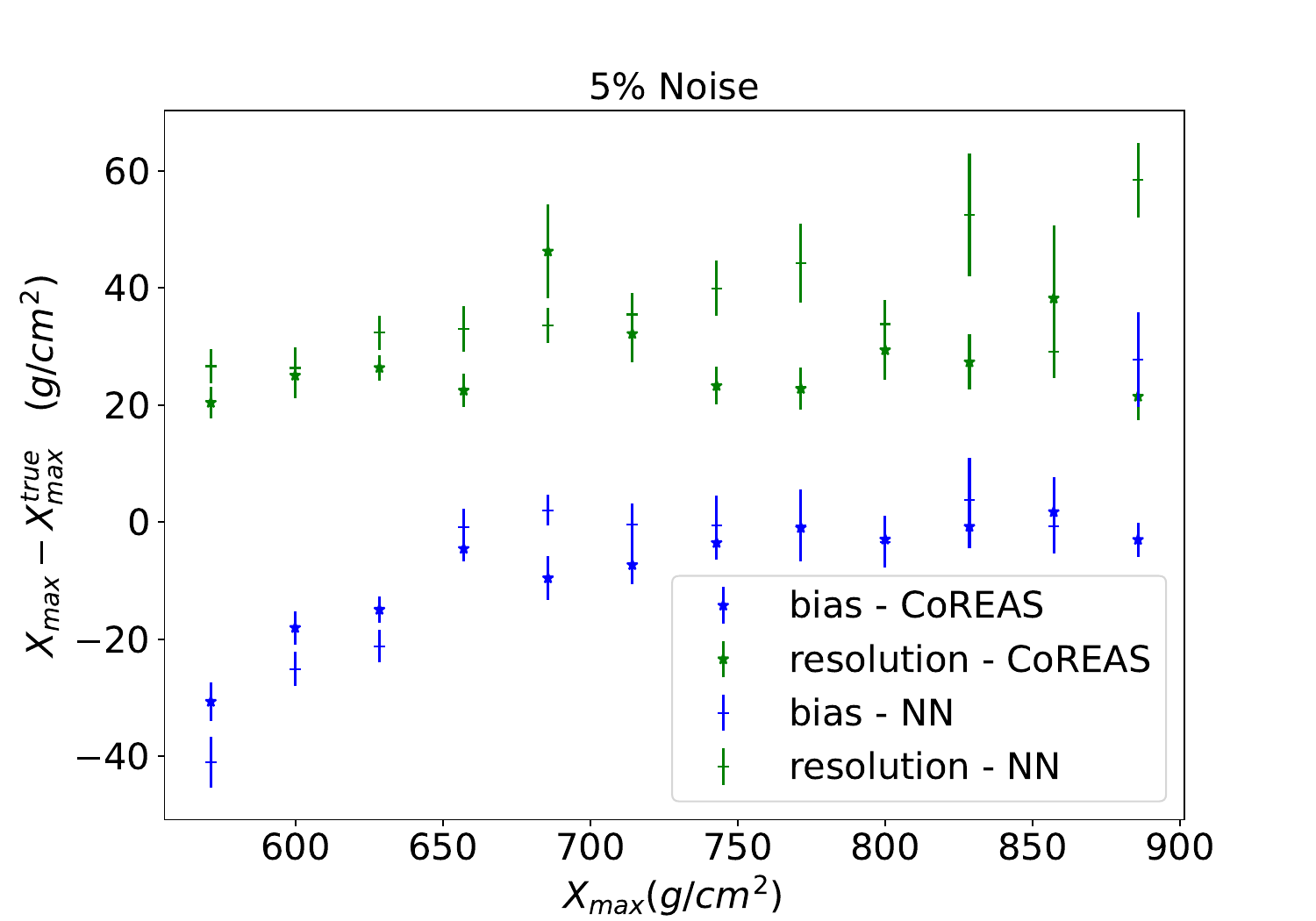}
    \includegraphics[width=0.45\textwidth]{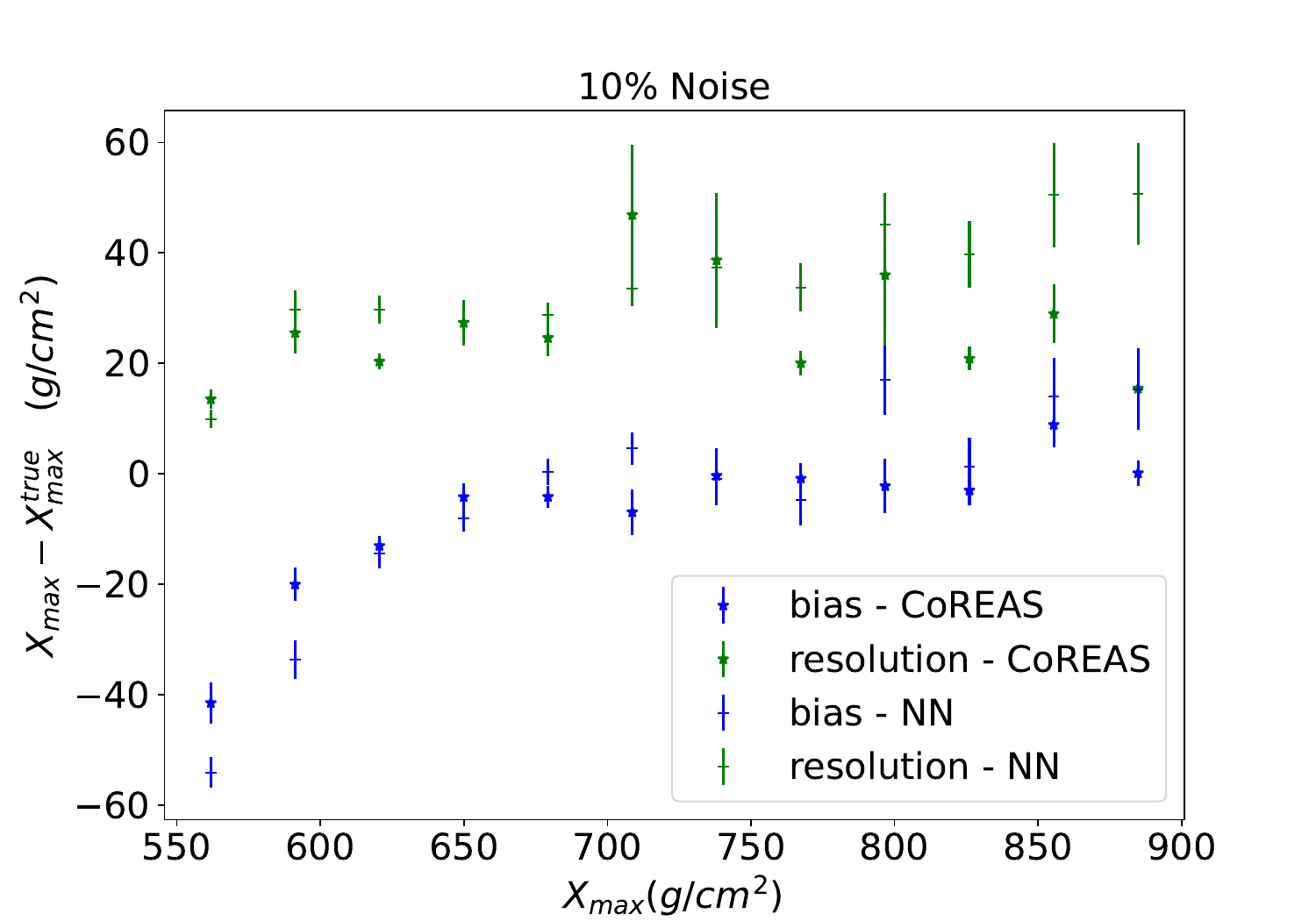}
    \caption{In this Figure, on the left, you see the $\xmax$ reconstruction procedure's bias and resolution for the (N=2000) simulations done using CoREAS and the neural network compared for 5\% Gaussian Noise. On the right we see the same for 10\% Gaussian Noise. As expected, there is a bias in the lower $\xmax$ region as result of uneven sampling on either side of the parabola fit. (also see, \Cref{fig:xmax_reco_coreas})}
    \label{fig:bias-resolution}
\end{figure*}
To benchmark the performance of using the neural network as a replacement for per-event CoREAS simulations in such an $\xmax$ reconstruction approach, we study the bias and resolution of the $\xmax$ reconstruction procedure for varying noise levels.
We also study the variation of the bias and resolution of the $\xmax$ reconstruction procedure with respect to varying $\xmax$ in \Cref{fig:bias-resolution}. 
But apart from expected bias at low $\xmax$, we can see that both the bias and resolution for the simulations are similar between neural network predictions and per-event CoREAS simulations, making neural networks a viable simulation tool for simulating radio pulses.
The total bias and resolution for all the events, irrespective of $\xmax$ is given in \Cref{tab:bias_res_results_gaussian_noise}. 
The neural network can be further optimized to improve these, but that is left for the future, since the current model's purposes can be broader than this $\xmax$ reconstruction. Beyond $\xmax$ of \qty{1000}{\gram\per\square\centi\metre}, the network performance falls drastically, because of the imbalanced input dataset. It is left to future work to improve this by either resampling the data or using a more balanced simulation training set distributed evenly across the phase space. 
\begin{table}[h]
    \centering
    \begin{tabular}{|c|c|c|c|c|}
        \hline
         &CoREAS - Bias& CoREAS - Res. & NN - Bias & NN - Res.\\
        \hline
         5\% Noise &  - 8.35 & 31.63 & -4.98 & 39.59\\
        \hline
         10\% Noise &  - 5.95 & 30.73 & -3.34 & 37.49\\
        \hline
    \end{tabular}
    \caption{The total bias and resolution for the $\xmax$ reconstruction procedure for using both methods. We can see that the reconstruction resolution is quite similar and thus the neural network can be used for simulations. The values are given in units of $\xmax$ (\qty{}{\gram\per\square\centi\metre}).}
    \label{tab:bias_res_results_gaussian_noise}
\end{table}

\section{Conclusion}
In this work, we presented a neural network that can be used to perform simulations of radio emissions from extensive air showers.
It provides a massive performance boost, as pulse predictions using neural networks can be performed almost instantly compared to traditional simulations.
Additionally, we also saw that the $\xmax$ reconstruction performance of the neural network is on par with using the CoREAS simulations.
The network is also shallow, and has a low memory footprint, so it could be easily run on FPGAs facilitating very quick simulations for triggering purposes \cite{fpga_nn}.
Also, neural networks provide the advantage of being differentiable, and thus can be effectively used with other frameworks such as Information Field Theory 
\cite{keito_ift}.
While this current work is a proof of concept, the network can be extended with additional input parameters, such as the detailed longitudinal profile of the entire Extensive Air Shower and thus can be used along with other frameworks to reconstruct the longitudinal profile from measured signals.
The trained model can also be used to fine-tune for other experiments to generate a model with smaller datasets.
Thus, the developed setup and the model can serve as a valuable contribution to the study of radio emission and the detection of cosmic rays using radio arrays. 


\let\oldbibliography\thebibliography
\renewcommand{\thebibliography}[1]{%
  \oldbibliography{#1}%
  \setlength{\itemsep}{1pt}%
}

{\footnotesize
\bibliographystyle{jhepnotitle}
\bibliography{references}
}

%
%
%
\end{document}